\documentclass[11pt]{article}

\usepackage[utf8]{inputenc}
\usepackage[T1]{fontenc}
\usepackage{graphicx}
\usepackage{longtable}
\usepackage{wrapfig}
\usepackage{rotating}
\usepackage[normalem]{ulem}
\usepackage{amsmath}
\usepackage{amssymb}
\usepackage{capt-of}
\usepackage{hyperref}
\usepackage[citestyle=authoryear-icomp,bibstyle=authoryear, hyperref=true,backref=true,maxcitenames=3,url=true,backend=bibtex]{biblatex}
\usepackage{multirow}
\usepackage{amsmath}
\usepackage{morefloats}
\usepackage{setspace}
\usepackage{graphicx}
\usepackage{float}
\usepackage{changepage}
\hypersetup{hidelinks}
\author{Matt Brigida\thanks{Chief Economist, Algorand Foundation (matthew.brigida@algorand.foundation) \& Department of Accounting and Finance, SUNY Polytechnic Institute, Utica, NY, USA}}
\date{\today}
\title{Crypto Pricing with Hidden Factors\thanks{I thank the editor and anonymous reviewers for helpful comments and suggestions that improved the paper.}}
\hypersetup{
 pdfauthor={Matt Brigida\thanks{Chief Economist, Algorand Foundation (matthew.brigida@algorand.foundation) \&  Department of Accounting and Finance, SUNY Polytechnic Institute, Utica, NY, USA}},
 pdftitle={Crypto Pricing with Hidden Factors},
 pdfkeywords={},
 pdfsubject={},
 pdfcreator={},
 pdflang={English}}
\begin{document}

\maketitle
\begin{abstract}
We estimate risk premia in the cross-section of cryptocurrency returns using the Giglio–Xiu (2021) three-pass approach, allowing for omitted latent factors alongside observed stock-market and crypto-market factors. Using weekly data on a broad universe of large cryptocurrencies, we find that crypto expected returns load on both crypto-specific factors and selected equity-industry factors associated with technology and profitability, consistent with increased integration between crypto and traditional markets.  In addition, we study non-tradable state variables capturing investor sentiment (Fear \& Greed), speculative rotation (Altcoin Season Index), and security shocks (hacked value scaled by market capitalization), which are new to the literature.  Relative to conventional Fama–MacBeth estimates, the latent-factor approach yields materially different premia for key factors, highlighting the importance of controlling for unobserved risks in crypto asset pricing.  Our findings should be interpreted as exploratory evidence from a short but institutionally important sample period.
\end{abstract}
\vspace*{1cm}



\footnotesize\noindent
Accepted manuscript. Accepted for publication in \emph{Finance Research Letters}.
This manuscript version is made available under the Creative Commons
Attribution-NonCommercial-NoDerivatives 4.0 International License (CC BY-NC-ND 4.0):
\url{https://creativecommons.org/licenses/by-nc-nd/4.0/}

\clearpage

Cryptocurrency markets have evolved rapidly alongside technological innovation, regulatory change, and growing interaction with traditional financial institutions.  A central question is which risks are priced in the cross-section of crypto returns.

We estimate cross-sectional prices of risk for crypto, stock-market, industry, and non-tradable state variables.  Prior work emphasizes crypto-native factors such as market, size, and momentum \cite{Liu_2020,liu2022common}.  We ask whether those patterns now coexist with broader equity-linked pricing forces.

In related research, \cite{Borri_2022} estimate crypto and stock-factor prices of risk using the \cite{Giglio_2021} latent factor model and find evidence that macro risk is priced in cryptocurrencies.  \cite{foley2022expected} and \cite{almeida2024risk} estimate Bitcoin-specific risk premia of 66--80\% per annum using option-implied and pricing-kernel approaches, confirming that investors demand substantial compensation for crypto risk.

Our 105-week sample from January 2023 through December 2024 captures the post-FTX recovery and the emergence of spot Bitcoin ETFs, so the estimates should be interpreted as exploratory evidence rather than definitive characterizations of long-run crypto risk premia.  The next section describes the data and factor construction.  Section 2 presents the factor-selection and latent-factor pricing results.  Section 3 concludes.
\section{Data}
\label{sec:orga7043db}

Our cryptocurrency data set is comprised of weekly returns over a sample period from January 1, 2023 through December 31, 2024.  We gather returns for any non-stablecoin cryptocurrency that was in the top 100 cryptocurrencies by market cap at any point in our sample period.  This sample construction is designed to ensure our estimates are not affected by a survivorship bias.

Using this method our sample contains 253 unique cryptocurrencies, which comprises over 97\% of the total crypto market capitalization. We do not attempt a wider sample because of poor price discovery for extremely low market capitalization cryptocurrencies.  Cryptocurrency prices were gathered via the Coinmarketcap Application Programming Interface.  Data on the US dollar amount hacked was provided by DeFiLlama\footnote{\url{https://defillama.com/}}.  Bitcoin implied volatility data is from the CVX website\footnote{\url{https://thecvx.com/api/chart}}.

The Altcoin Season Index is from the CoinMarketCap website and measures the relative performance of altcoins relative to Bitcoin. All cryptocurrencies other than Bitcoin are referred to as \emph{altcoins}.  The Fear/Greed index is also from CoinMarketCap and is a measure of market sentiment.

Stock factor data was downloaded from Kenneth French's website\footnote{\url{http://mba.tuck.dartmouth.edu/pages/faculty/ken.french/data\_library.html}}.  Factor inclusion is motivated as follows.  Standard Fama-French factors (\(R_S\), \(SMB_S\), \(HML_S\), \(Mom_S\), \(RMW_S\), \(CMA_S\)) are included to test whether equity factor premia extend to crypto, consistent with increased institutional participation and market integration over the 2023--2024 sample period.  Industry factors are included for several economic reasons: \(Softw\) and \(Chips\) capture technology supply-chain linkages (computing resources underpin blockchain validation) and shared investor bases between technology and crypto; \(Banks\), \(Insur\), and \(Fin\) capture potential substitution between traditional financial products and crypto; and \(Gold\) provides an alternative-store-of-value comparison.  Non-tradable state variables are selected for specific reasons: the Fear/Greed index captures market-wide sentiment shifts; the Altcoin Season Index captures speculative rotation between Bitcoin and altcoins; Hacks scaled by market cap captures evolving security risk; and CVX captures crypto-specific implied volatility conditions.
\subsection{Crypto Factor Construction}
\label{sec:orga5f0910}

Below we describe how we construct each crypto factor. Factor construction follows the methods used by \cite{fama1993common}.  To construct crypto market returns, we compute the weekly value-weighted return on the aggregate crypto market using total market capitalization from CoinMarketCap\footnote{\url{https://coinmarketcap.com/}}, and subtract the risk-free rate to obtain excess returns.  We correct the market capitalization data for token-migration and tracking-inception artifacts that would otherwise mechanically generate uninvestable returns when newly tracked entities enter the small-cap portfolio. The Supplementary Appendix details the affected cases and corrections.

To construct the TVL factor we first scale TVL by market cap which affords a proportion of market cap which is locked.  Then, for a given week \(t\), we rank assets by TVL / Market Cap in week \(t-1\).  We then create a value-weighted long-short portfolio which buys the top 25\% of coins by TVL and sells the bottom 25\%.  The return on this portfolio over week \(t\) is the factor realization for week \(t\).  This high-minus-low TVL portfolio is denoted as \(TVL\) in tables below. Given evidence that TVL factor returns are spanned by the crypto market portfolio (\cite{BRIGIDA2025112673}), we orthogonalize TVL with respect to crypto market returns.

To calculate the momentum factor over week \(t\) we calculate the cumulative return for each cryptocurrency over weeks \(t-5\) to \(t-1\).  We then sort the cryptocurrencies and create a long-short portfolio which buys the currencies in the top 25\% of cumulative return, and sells the currencies in the bottom 25\%.  The return on this portfolio over week \(t\) is our crypto momentum factor for week \(t\).  We repeat this procedure for all weeks in our sample.

To construct the SMB factor for week \(t\), we sort all cryptocurrencies by market capitalization in week \(t-1\). We then create a value-weighted long-short portfolio that buys the bottom 25\% of coins by size and shorts the top 25\%. The return on this portfolio over week \(t\), less the risk-free rate, is the factor realization for week \(t\).

Hacks is scaled by market cap, and the Fear and Greed and Altseason indices are converted to percent change.  The CVX volatility index is maintained in levels.  We then convert each non-tradeable factor into its residual component via an AR(1) model.
\subsection{Descriptive Statistics}
\label{sec:org91306d4}

Descriptive statistics for all factors are in Table 1 below.  Crypto excess market returns average 1.43\% per week, substantially above the stock market's 0.36\% weekly return. Crypto market returns have over 3 times the standard deviation of stock market returns, and lower kurtosis (2.05 versus 6.01).

The crypto size long-short portfolio (buying the bottom 25\% of coins by market capitalization and shorting the top 25\%) has a mean weekly return of 1.21\% after correcting for token migration artifacts (see Section 1.1).  The minimum weekly return is -14.46\% and the standard deviation is 6.75\%.  The positive average return indicates that small-cap crypto slightly outperformed large-cap over this sample.

The crypto momentum portfolio has a lower weekly mean return than the crypto market (0.49\%) and slightly higher standard deviation.

\begin{table}[htbp]
\caption{\label{tab:orgcc9cf3f}Factor Descriptive Statistics.  105 weekly observations, Jan. 1, 2023 to Dec. 31, 2024.  Mean, Std, Min, and Max are weekly values expressed in percent.  Kurtosis uses Fisher’s definition (normal distribution = 0).  \(R_C\) and \(R_S\) are excess crypto and stock market returns.  \(SMB_C\) values reflect corrected entity-linked data (see Section 1.1).  Non-tradable factors are AR(1) residuals of percent changes (\(Altseason\), \(Fear/Greed\)) or levels (\(CVX\), \(Hacks\)).}
\centering
\begin{tabular}{lrrrrrr}
\hline
Factor & Mean & Std & Min & Max & Skewness & Kurtosis\\
\hline
\multicolumn{7}{l}{\textit{Panel A: Crypto Factors}}\\
\hline
\(R_C\) & 1.43 & 6.35 & -15.42 & 18.69 & -0.45 & 2.05\\
\(SMB_C\) & 1.21 & 6.75 & -14.46 & 24.59 & 0.74 & 1.62\\
\(Mom_C\) & 0.49 & 7.37 & -34.48 & 19.18 & -0.64 & 4.03\\
\(TVL\) & -0.11 & 5.70 & -13.99 & 19.94 & 0.84 & 2.28\\
\hline
\multicolumn{7}{l}{\textit{Panel B: Stock Market Factors}}\\
\hline
\(R_S\) & 0.36 & 1.91 & -5.75 & 5.66 & -0.18 & 6.01\\
\(SMB_S\) & -0.13 & 1.72 & -4.03 & 5.26 & -0.04 & 1.13\\
\(HML_S\) & -0.18 & 1.45 & -4.43 & 3.46 & 0.44 & 3.72\\
\(Mom_S\) & 0.05 & 1.81 & -5.33 & 4.62 & -1.27 & 8.80\\
\(RMW_S\) & 0.05 & 1.08 & -2.09 & 2.65 & 0.07 & -0.49\\
\(CMA_S\) & -0.23 & 0.94 & -2.65 & 2.88 & 0.25 & 0.64\\
\hline
\multicolumn{7}{l}{\textit{Panel C: Non-Tradable Factors}}\\
\hline
\(Hacks\) & 0.00 & 0.00 & 0.00 & 0.02 & 3.25 & 11.26\\
\(Altseason\) & 2.99 & 23.45 & -40.98 & 84.85 & 0.85 & 1.19\\
\(Fear/Greed\) & 0.18 & 1.85 & -5.28 & 4.88 & -0.24 & 0.48\\
\(CVX\) & 0.01 & 0.14 & -0.30 & 0.46 & 0.52 & 0.40\\
\hline
\end{tabular}
\end{table}
\section{Method and Results}
\label{sec:org5974ca4}

Theory on cryptocurrency pricing is still in its infancy, and so there is little guidance on factors which will affect prices.  Given the likelihood of unobserved factors when building factor models of cryptocurrencies, we use the \cite{Giglio_2021} latent factor model.  We estimate prices of risk using the Giglio and Xiu (2021) three-pass estimator, which recovers latent factors from the covariance structure of returns and maps observed factor realizations into prices of risk while allowing for omitted common risks.  We estimate seven latent factors selected by the Bai-Ng information criteria and compute inference using a moving-block bootstrap with block length eight and 1,000 replications.  The Supplementary Appendix provides implementation details for the unbalanced panel, latent-factor estimation, and bootstrap procedure.\\

\noindent \emph{Fama-Macbeth} \\

We use a standard \cite{Fama_1973} estimation procedure, and \cite{Shanken_1992} adjusted standard errors.  For an exposition of the method see \cite{Cochrane2005AssetPricing}.
\subsection{Factor Selection}
\label{sec:orgdd2ea9d}

To provide a principled basis for factor inclusion, we estimate penalized cross-sectional regressions using elastic net regularization.  The elastic net combines the ridge penalty (appropriate for the multicollinear beta structure in cross-sectional asset pricing, as employed by \cite{Giglio_2021}) with the LASSO penalty, which allows the data to exclude irrelevant factors.  The methodology is a penalized analogue of the Fama-MacBeth cross-sectional step.  We first estimate time-series betas of each asset on all candidate factors via OLS, then regress mean returns on estimated betas with cross-validated elastic net penalty selection.  The elastic net regularization path, which is the sequence of penalty values at which each factor first receives a non-zero coefficient, ranks factors by their marginal contribution to cross-sectional pricing.\footnote{\cite{cheng2016shrinkage} propose a shrinkage estimator for factor models with structural instabilities.  Our 105-week sample limits formal break detection, so we assess factor stability via bootstrap resampling \parencite{meinshausen2010stability} rather than full structural-break estimation.  Results are qualitatively similar when using LASSO rather than elastic net penalization.}

\begin{table}[htbp]
\caption{\label{tab:org27497cd}Factor Selection: Elastic Net Regularization Path and Stability Selection.  Entry rank indicates the order in which factors receive non-zero coefficients as the penalty decreases from its maximum.  The elastic net combines ridge and LASSO penalties, handling multicollinearity among estimated betas while allowing factor exclusion.  Stability frequency reports the percentage of 500 block-bootstrap samples (block size 8) in which each factor is selected by cross-validated LASSO \parencite{meinshausen2010stability}.  Only the first 10 factors to enter are shown; remaining factors enter subsequently with stability frequencies between 9\% and 23\%.}
\centering
\begin{tabular}{rlr}
\hline
Entry Rank & Factor & Stability Freq. (\%)\\
\hline
1 & Util & 23.0\\
2 & RMW & 18.6\\
3 & Mkt-RF & 17.4\\
4 & Softw & 21.8\\
5 & SMB (stock) & 15.4\\
6 & TVL & 13.2\\
7 & SMB (crypto) & 28.0\\
8 & RlEst & 11.0\\
9 & Mom (stock) & 9.8\\
10 & Gold & 13.0\\
\hline
\end{tabular}
\end{table}

The factors entering earliest under penalization are those with the strongest marginal contribution to cross-sectional pricing and they overlap substantially with the factors that carry significant prices of risk in the three-pass estimation (Table 3).  The stock market profitability factor (\(RMW_S\)), stock market returns (\(R_S\)), and the Software industry portfolio all enter within the first four positions, consistent with their significant prices of risk in the latent factor model.  The crypto size factor enters seventh, consistent with its role in cross-sectional pricing though not as dominant as the equity-linked factors under penalization.

Under cross-validated penalty selection, neither elastic net nor LASSO retains any factor.  This is consistent with the absence of a single dominant predictor and with the structure of many small premia factors which is the motivation of the ridge approach in \cite{Giglio_2021}.  To assess robustness, we follow the stability selection framework of \cite{meinshausen2010stability}, repeating the LASSO selection on 500 block-bootstrap samples.  Selection frequencies range from 9\% to 28\%, with no clearly dominant factor, further supporting the diffuse risk structure identified by the latent factor model.

Notably, the Utilities factor enters first in the regularization path but is insignificant in both the three-pass and Fama-MacBeth estimations (Table 3), indicating that its cross-sectional sorting power reflects correlation with latent common factors rather than an independent price of risk.  This illustrates the value of the latent factor correction for distinguishing genuine priced risk from spurious cross-sectional associations.
\subsection{Latent Factor Model Results}
\label{sec:org1e39984}

Results from estimating the \cite{Giglio_2021} model over the 2023--2024 sample are in Table 3.  For comparison, we also include Fama-MacBeth estimates.  The factors identified as significant by the latent factor model are among those entering earliest in the elastic net regularization path (Table 2), providing complementary evidence that the results are not driven solely by an ad hoc factor list.

The latent factor model provides suggestive evidence of a 0.471\% price of risk for crypto market returns (p=0.062), which corresponds to an annualized premium of 24.5\%.  The Fama-MacBeth estimate of 0.116\% per week (annualized 6.03\%, p<0.001) is strongly significant, and the divergence across methods highlights the importance of controlling for latent factors.  \cite{Borri_2022} estimate a similar annualized crypto market premium of 26\% using a similar methodology.

Both methods estimate a significantly positive price of risk for the crypto size (\(SMB_C\)) factor.  The GX estimate is 0.430\% (p=0.008), corresponding to an annualized premium of approximately 22.4\%, while the FM estimate is 0.080\% (p<0.001).  The positive size premium is consistent with small-cap crypto carrying higher expected returns over this sample.

The latent model provides suggestive evidence of a positive price of risk for the Software stock portfolio (p=0.056), overall stock market returns (p=0.066), and a negative price of risk for the stock market profitability factor \(RMW_S\) (p=0.068).  The Fama-MacBeth results similarly price Software and additionally the Finance and Insurance sectors.

There is suggestive evidence of a negative price of risk associated with innovations in the Fear/Greed index (p=0.058).  Altseason innovations are insignificant in the latent factor model but significant in Fama-MacBeth, consistent with proxying for latent common factors.  Hack innovations are indistinguishable from zero.  The Hacks variable exhibits near-zero variation over the sample because the major Bybit hack (USD 1.5 billion, February 21, 2025) occurred after our sample ends, limiting power to detect a hack-related price of risk.

The Fama-MacBeth method provides weak evidence that TVL is priced (p=0.068), while the \cite{Giglio_2021} latent factor model finds TVL’s incremental price of risk is not distinguishable from zero at conventional levels.  This is consistent with TVL proxying for broad latent risk rather than carrying an independent price of risk.

\begin{table}[htbp]
\caption{\label{tab:orgc7eb507}Giglio and Xiu Latent Factor and Fama-MacBeth Model Results.  The weekly sample ranges from 2023-01-02 to 2024-12-31 for all cryptocurrencies that were in the top 100 by market capitalization at any point in the sample.  \(R_C\) and \(R_S\) are excess crypto and stock market returns.  \(SMB_S\), \(HML_S\), and \(Mom_S\) are weekly Fama-French factors.  \(Mom_C\) and \(SMB_C\) are long-short crypto momentum and small-minus-big portfolios.  \(Hacks\) is the proportion of crypto market cap hacked each week.  \(Altseason\) and Fear/Greed are the Altcoin Season and Fear/Greed indices.  Asterisks denote significance at the 1\%, 5\%, and 10\% levels, respectively.}
\centering
\begin{tabular}{lrrllr}
\hline
 & Giglio-Xiu Latent & p-value &  & Fama-MacBeth & p-value\\
\hline
\(SMB_C\) & 0.430\(^{***}\) & 0.008 &  & 0.080\(^{***}\) & 0.000\\
Softw & 0.071\(^{*}\) & 0.056 &  & 0.028\(^{***}\) & 0.002\\
\(Fear\ and\ Greed\) & -0.051\(^{*}\) & 0.058 &  & -0.008\(^{**}\) & 0.035\\
\(R_C\) & 0.471\(^{*}\) & 0.062 &  & 0.116\(^{***}\) & 0.000\\
\(R_S\) & 0.064\(^{*}\) & 0.066 &  & 0.006\(^{*}\) & 0.090\\
\(RMW_S\) & -0.033\(^{*}\) & 0.068 &  & -0.006\(^{**}\) & 0.043\\
\(TVL\) & 0.257 & 0.128 &  & 0.026\(^{*}\) & 0.068\\
\(SMB_S\) & 0.046 & 0.152 &  & 0.005 & 0.173\\
\(Altseason\) & 0.475 & 0.240 &  & 0.054\(^{**}\) & 0.023\\
Fin & 0.033 & 0.366 &  & 0.010\(^{**}\) & 0.037\\
Banks & 0.034 & 0.366 &  & 0.012\(^{*}\) & 0.059\\
Chips & 0.033 & 0.404 &  & 0.012 & 0.172\\
Insur & 0.021 & 0.424 &  & 0.017\(^{***}\) & 0.009\\
\(Mom_S\) & 0.024 & 0.523 &  & 0.001 & 0.857\\
\(HML_S\) & 0.020 & 0.601 &  & -0.000 & 0.862\\
\(CMA_S\) & -0.011 & 0.667 &  & -0.002 & 0.319\\
Util & -0.021 & 0.669 &  & 0.001 & 0.795\\
RlEst & 0.006 & 0.863 &  & 0.007 & 0.320\\
\(Mom_C\) & -0.087 & 0.891 &  & -0.003 & 0.842\\
Gold & -0.007 & 0.879 &  & -0.012 & 0.357\\
CVX & 0.003 & 0.899 &  & 0.000 & 0.967\\
\(Hacks\) & 0.000 & 0.931 &  & 0.000 & 0.233\\
\hline
\hline
\end{tabular}
\end{table}
\subsection{Robustness Checks}
\label{sec:org9365318}

To assess the sensitivity of our latent factor model results, we examine six robustness dimensions: varying the number of latent factors, block-bootstrap length, extreme crypto-market weeks, a USD 500 million market-capitalization screen, equal-weighted construction of the crypto size factor, and portfolio breakpoints.  Table 4 reports estimates for the two most consistently significant factors across the first five checks.  Breakpoint robustness, reported in Supplementary Appendix Table A1, similarly shows that the size result is not driven by the 25/75 portfolio-sort choice.

\begin{table}[htbp]
\caption{\label{tab:org089dcbb}Robustness of Latent Factor Model Estimates.  Baseline specification uses \(K=7\) latent factors chosen by Bai-Ng IC and block-bootstrap length \(b=8\).  "Trimmed" excludes the top and bottom 5\% of crypto market return weeks.  "Screened" imposes a USD 500 million large-cap screen on included assets.  "EW" replaces value-weighted with equal-weighted returns in the \(SMB_C\) sort legs.  Asterisks denote significance as in Table 3.}
\centering
\begin{tabular}{lllllllrl}
\hline
Factor & Baseline & K=5 & K=9 & b=4 & b=12 & Trimmed & Screened & EW\\
\hline
\(SMB_C\) & 0.430\(^{***}\) & 0.427\(^{**}\) & 0.549\(^{***}\) & 0.430\(^{***}\) & 0.430\(^{**}\) & 0.713\(^{***}\) & 0.150 & 0.981\(^{***}\)\\
Softw & 0.071\(^{*}\) & 0.067\(^{*}\) & 0.148\(^{*}\) & 0.071\(^{**}\) & 0.071\(^{*}\) & 0.139\(^{*}\) & 0.142 & 0.071\(^{*}\)\\
\hline
\end{tabular}
\end{table}

The positive crypto size premium is robust across these checks except under the large-cap screen, consistent with the premium being concentrated among smaller-cap assets.  The equal-weighted estimate exceeds the value-weighted baseline, indicating that the result is not driven by a few large coins dominating the portfolio legs.
\section{Conclusion}
\label{sec:org44bd132}

Across 2023–2024, exploratory evidence from our 105-week sample is suggestive of a view of crypto as increasingly intertwined with traditional equity risks rather than as a fully segmented asset class.  In addition to crypto-native exposures (market and size), the latent-factor estimates indicate suggestive prices of risk for the Software industry portfolio, overall stock market returns, and the Robust Minus Weak (\(RMW_S\)) stock profitability factor.  This pattern is consistent with shared investor bases and common “risk-on/risk-off” channels linking technology equities and crypto assets.

We also estimate prices of risk associated with innovations in the amount of crypto hacked, the altseason index, and the crypto Fear/Greed index.  We find no evidence that hack innovations carry a nonzero price of risk, limited suggestive evidence that altseason index innovations are priced, and suggestive evidence that Fear/Greed innovations have explanatory power for expected returns.  The small variation in Hacks over the sample limits inferences about security risk pricing.

Lastly, we shed light on the impact of crypto factors.  The size factor carries a positive 22.4\% annualized price of risk, indicating that small-cap crypto assets earn higher expected returns, consistent with a crypto size premium analogous to the stock market size effect.  Our results with respect to TVL complement previous research and find TVL does not carry an independent price of risk after latent-factor adjustment.  Future research should test whether these relationships persist in longer samples, across different market regimes, and under alternative liquidity and market-microstructure assumptions.
\section{Supplementary Appendix}
\label{sec:orgffab0e5}

\subsection{Giglio-Xiu Implementation Details}
\label{sec:orgeb1547b}

The Giglio-Xiu estimator extends the classical Fama-MacBeth framework by explicitly incorporating unobservable common factors.  Let \(r_{i,t}\) denote the excess return on cryptocurrency \(i=1,\dots,N_t\) in week \(t=1,\dots,T\), where the panel is unbalanced and \(N_t\) varies by week:
\begin{equation}
r_{i,t} \;=\; \alpha_i \;+\; \beta_i^{\top} u_t \;+\; \varepsilon_{i,t},
\end{equation}
where \(u_t \in \mathbb{R}^{K}\) are latent common factors and \(\beta_i\) are asset-specific loadings.

\noindent \emph{Pass 1:} We estimate \(u_t\) by principal components on the \(T\times N\) panel of excess returns, yielding \(\widehat{u}_t\).  Let \(g_t\in\mathbb{R}^{L}\) be the vector of observed factor realizations.

\noindent \emph{Pass 2:} We estimate the mapping from latent factors to observed factors using
\begin{equation}
g_t \;=\; a \;+\; \Lambda \widehat{u}_t \;+\; e_t,
\end{equation}
which yields \(\widehat{\Lambda}\in\mathbb{R}^{L\times K}\).

\noindent \emph{Pass 3:} We estimate the latent factor prices of risk \(\gamma\in\mathbb{R}^{K}\) from cross-sectional regressions and map them into prices of risk for the observed factors,
\begin{equation}
\widehat{\lambda}_g \;=\; \widehat{\Lambda}\,\widehat{\gamma} \;\in\; \mathbb{R}^{L}.
\end{equation}

Because the cryptocurrency return panel is unbalanced, we repeatedly fill in missing weekly returns with values implied by a \(K\)-factor structure while keeping observed returns unchanged until the completed matrix stabilizes, and then apply standard PCA to that filled-in matrix.  Inference is based on a moving block bootstrap over the time index.  We resample overlapping blocks of length \(b\) weeks with replacement, concatenate them until reaching sample length \(T\), recompute the full three-pass estimator in each replication, and form recentered two-sided bootstrap p-values as
\[
p_j =
\Pr^{*}\!\left(
\left|\lambda_{j}^{*}-\widehat{\lambda}_{j}\right|
\ge
\left|\widehat{\lambda}_{j}\right|
\right).
\]
In the baseline specification, \(K=7\) is chosen by the Bai-Ng information criteria, \(b=8\), and the number of bootstrap replications is \(B=1000\).
\subsection{Crypto Size-Factor Data Corrections}
\label{sec:orgebe6c44}

The affected cases are BitDAO/Mantle, dYdX, and Bittensor. We entity-link BitDAO with Mantle, sum the Ethereum and Cosmos based dYdX entries to reflect total project value, and exclude Bittensor because CoinMarketCap begins tracking an already-existing token without a predecessor series that can be linked.
\subsection{Breakpoint Robustness}
\label{sec:org6d100c7}

\renewcommand{\thetable}{A\arabic{table}}
\setcounter{table}{0}

\begin{table}[htbp]
\caption{\label{tab:orgda4ee2e}Breakpoint Robustness of Crypto Factor Estimates (\(K=7\), \(b=8\), \(B=1000\)).  Each specification replaces the three baseline crypto factors with versions constructed using the indicated portfolio breakpoint applied to the top-100 CMC rank universe (stablecoins excluded).  The 25/75 column reproduces the baseline estimates from Table 3.  Asterisks denote significance as in Table 3.}
\centering
\begin{tabular}{llrrrrr}
\hline
Factor & 20/80 & p-value & 25/75 & p-value & 30/70 & p-value\\
\hline
\(SMB_C\) & 0.413\(^{***}\) & 0.008 & 0.430\(^{***}\) & 0.008 & 0.325\(^{***}\) & 0.010\\
\(Mom_C\) & -0.094 & 0.941 & -0.087 & 0.891 & -0.096 & 0.557\\
\(TVL\) & 0.402\(^{*}\) & 0.082 & 0.257 & 0.128 & 0.156 & 0.178\\
\hline
\end{tabular}
\end{table}

The positive price of risk for \(SMB_C\) is robust to the portfolio breakpoint choice: estimates range from 0.325 to 0.430 and are significant at the 1\% level across all three specifications.  \(Mom_C\) remains statistically indistinguishable from zero at all breakpoints.  The TVL price of risk is borderline at the 20/80 sort (p=0.082) but insignificant at 25/75 and 30/70, consistent with the baseline finding that TVL carries no independent price of risk after latent-factor adjustment.

\clearpage

\clearpage

\printbibliography
\end{document}